%% file: main.tex
\documentclass[conference]{IEEEtran}
\IEEEoverridecommandlockouts
\usepackage{cite}
\usepackage{amsmath,amssymb,amsfonts}
\usepackage{algorithmic}
\usepackage{graphicx}
\usepackage{textcomp}
\usepackage{xcolor}
\usepackage{url}
\def\BibTeX{{\rm B\kern-.05em{\sc i\kern-.025em b}\kern-.08em
    T\kern-.1667em\lower.7ex\hbox{E}\kern-.125emX}}

\usepackage{tcolorbox}
\usepackage{minted}
\usepackage{flushend}
\usepackage{cite}
\usepackage{amsmath,amssymb,amsfonts}
\usepackage{algorithmic}
\usepackage{graphicx}
\usepackage{textcomp}
\usepackage{hyperref}
\usepackage{xcolor}
\usepackage{color}
\usepackage{booktabs}
\usepackage{xspace}
\usepackage{graphicx}
\usepackage{multirow}
\usepackage[framemethod=TikZ]{mdframed}
\usepackage{textcomp}
\usepackage{booktabs}
\usepackage{colortbl}
\usepackage{ragged2e}
\usepackage{courier}
\usepackage{textcomp} 
\usepackage{listings}
\usepackage{varwidth}
\usepackage{subcaption}

\def\BibTeX{{\rm B\kern-.05em{\sc i\kern-.025em b}\kern-.08em
    T\kern-.1667em\lower.7ex\hbox{E}\kern-.125emX}}

\usepackage[normalem]{ulem}

\newboolean{showcomments}
\setboolean{showcomments}{true}
\ifthenelse{\boolean{showcomments}}
	{\newcommand{\nb}[3]{
		{\colorbox{#2}{\bfseries\sffamily\scriptsize\textcolor{white}{#1}}}
		{\textcolor{#2}{\sf\small$\blacktriangleright$\textit{#3}$\blacktriangleleft$}}}
	 }
	{\newcommand{\nb}[3]{}
	 }

\usepackage{xcolor}
\usepackage{textcomp}
\usepackage[T1]{fontenc}
\usepackage{inconsolata}

\newcommand{\nadel}[1]{}  

\newcounter{rmd}

\begin{document}

\title{\textit{Not Only for Developers}: Exploring Plugin Maintenance for Knowledge-Centric Communities}

 \author{
 \IEEEauthorblockN{
 Giovanni Rosa\IEEEauthorrefmark{1},
 David Moreno-Lumbreras\IEEEauthorrefmark{1},
 Raula Gaikovina Kula\IEEEauthorrefmark{2}
 }
 \IEEEauthorblockA{\IEEEauthorrefmark{1}Universidad Rey Juan Carlos, Spain}
 \IEEEauthorblockA{\IEEEauthorrefmark{2}The University of Osaka, Japan}
 }

\maketitle

\begin{abstract}

The adoption of third-party libraries has become integral to modern software development, leading to large ecosystems such as PyPI, NPM, and Maven, where contributors typically share the technical expertise to sustain extensions. In communities that are not exclusively composed of developers, however, maintaining plugin ecosystems can present different challenges. In this early results paper, we study \textit{Obsidian}, a knowledge-centric platform whose community—focused on writing, organization, and creativity—has built a substantial plugin ecosystem despite not being developer-centric.
We investigate what kinds of plugins exist within this hybrid ecosystem and establish a foundation for understanding how they are maintained. Using repository mining and LLM-based topic modeling on a representative sample of 396 plugins, we identify six topics related to knowledge management and tooling, which is (i) dynamic editing and organization, (ii) interface and layouts, (iii) creative writing and productivity, (iv) knowledge sync solutions, (v) linking and script tools, and (vi) workflow enhancements tools. 
Furthermore, analysis of the Pull Requests from these plugins show that much software evolution has been performed on these ecosystem.
These findings suggest that even in mixed communities, plugin ecosystems can develop recognizable engineering structures, motivating future work that highlight three different research directions with six research questions related to the health and sustainability of these non-developer ecosystems.

\end{abstract}


\section{Introduction}

Many researchers and practitioners are likely to assume that software ecosystems are environments created \emph{by developers and for developers}.  
Platforms such as Visual Studio Code, NPM, PyPI, and JetBrains IDEs represent this classical view: ecosystems in which plugins and extensions support programming workflows, accelerate development, enable automation, and integrate specialized engineering tools~\cite{edirimannage2024developersvictimscomprehensive}.  
Within this paradigm, ecosystem evolution, maintainability, and governance have been studied primarily through the lens of software engineering practices, with the implicit assumption that both producers and consumers share technical expertise.

However, modern software ecosystems increasingly support large and diverse communities whose goals extend far beyond programming.  
In these environments, plugins do not merely enhance developers’ productivity—they expand the cognitive and creative capabilities of users who may have little or no background in software development.  
This shift raises new questions about how such ecosystems form, how they evolve, and how developers sustain them when the primary beneficiaries are not developers but a broader knowledge-centric community.

Another motivation is how there is need for a richer understanding of hybrid software ecosystems that extend beyond developer-centric ecosystems, especially in the era of Generative Artificial Intelligence (GenAI).  
As large language models further reduce the barriers to customization and automation~\cite{Amershi2019, Marron2024}, the boundary between coding tasks and thinking tasks becomes increasingly blurred.  
Platforms like Obsidian illustrate this transition: developers provide plugins that enable users to construct, navigate, and reason about knowledge through software artifacts. 

\textit{Obsidian}---a Markdown-based knowledge management application---is a compelling example of this emerging class of hybrid software ecosystems.  
Originally designed as a note-taking environment, Obsidian has grown into a modular, community-driven platform enhanced by thousands of plugins hosted on GitHub.  
These plugins span a wide spectrum of functionality: from task and calendar management, to AI-powered writing assistants, to visualization tools, diagram editors, music notation utilities, and flashcard generation.  
Crucially, these plugins do not extend a programming environment.  
They extend a set of \emph{knowledge-centric workflows}: writing, structuring information, planning tasks, externalizing ideas, and building personal knowledge systems.

Open plugin ecosystems such as VS Code, Eclipse, Chrome, and WordPress have been extensively studied for their structure, evolution, and governance challenges. Prior work has examined risks such as malicious extensions~\cite{edirimannage2024developersvictimscomprehensive}, malware-related patterns in browser ecosystems~\cite{Hsu_2024}, and incompatibilities arising from high coupling in content-management platforms~\cite{linwordpress2023}. These studies focus on \emph{developer-centric} environments, whereas Obsidian extends a \emph{knowledge environment} in which plugins support cognitive rather than programming tasks. Parallel research on large language models (LLMs) has shown their usefulness in repository classification~\cite{zanartu2022automatically,sas2023gitranking}, workflow categorization~\cite{nguyen2024gavel}, and human–AI collaboration~\cite{Amershi2019}, while also addressing challenges of trust, usability, and evaluation in LLM-powered tools~\cite{Sergeyuk2024,Nghiem2024,Ahmed2025,Crupi2025}. Broader perspectives envision intelligent development environments in which AI augments reasoning rather than just code editing~\cite{Gonzalez-Barahona2024,Marron2024}. Complementary efforts on plugin classification and recommendation, such as systems for the Chrome Web Store~\cite{qin2021chrome} and GitHub Actions~\cite{nguyen2024gavel}, target software-production ecosystems; in contrast, our work focuses on \emph{knowledge-centric communities}, expanding the scope of modern software ecosystems toward tools that mediate reasoning, learning, and personal organization.

This inversion of roles—developers creating tools for a knowledge-centric community—poses fundamental questions for software engineering research.  
What kinds of software artifacts emerge in such hybrid ecosystems?  
How do developers reason about users whose goals are not programming tasks but thinking tasks?  
What kinds of development and maintenance practices are sustained when the primary audience is less likely to contribute technical implementations?  
And to what extent do these ecosystems resemble, or diverge from, the structural patterns observed in traditional developer-centric ecosystems, such as collaborative maintenance models and standard contribution flows?

Despite its rapid adoption and vibrant plugin ecosystem, Obsidian has received \emph{limited systematic attention} from the software engineering community, especially when compared to ecosystems such as VS Code or the Chrome Web Store~\cite{qin2021chrome}.  
This absence of formal analysis obscures fundamental questions: how the ecosystem is organized, which functionalities dominate, how specialized its extensions are, or how developers maintain them over time.  
Moreover, Obsidian sits at the intersection of software engineering, human–computer interaction, personal knowledge management, and cognitive augmentation—making it an ideal setting to explore emerging hybrid software ecosystems.

Motivated by these gaps, this research preview presents an early investigation of the Obsidian plugin ecosystem from a software engineering standpoint.  
We combine repository mining, clustering, and large language models (LLMs) to uncover the functional landscape of the ecosystem and to analyze the extent to which plugins are maintained.  
This study is exploratory by design: rather than delivering a complete evaluation, we aim to assess the feasibility and value of applying software engineering techniques to an ecosystem whose primary purpose is not software production.

Our investigation is guided by the following research questions:

\begin{itemize}
    \item \textit{\textbf{RQ1: What kinds of plugins exist within the Obsidian ecosystem?}}  
    This question is motivated by the need to understand how functionality is organized in a hybrid ecosystem where developers build plugins for a knowledge-centric community, rather than for programming workflows.

    \item \textit{\textbf{RQ2: To what extent are these plugins maintained with respect to Pull Requests to these plugins?}}  
    This question arises from the uncertainty about whether plugins used primarily by non-developers follow maintenance patterns similar to traditional developer-centric ecosystems.
\end{itemize}

Our research shows that such non developer ecosystems do evolve their software systems, which brings three different research directions with six research questions.
A complete reproduction package, including all scripts, processed data, and artifacts used in this preliminary analysis, is publicly available for inspection and reuse at {\url{https://doi.org/10.5281/zenodo.17632440}.

\newcommand{\bertopic}{\emph{BERTopic}\xspace}

\section{Envisioned Methodology: A Case of Obsidian}
The proposed methodology aims to construct an initial, data-driven taxonomy of the Obsidian plugin ecosystem composed of three main stages, namely (i) Data Collection and Preparation, (ii) Approach to answer RQ1: Topic Modeling and Labeling, and (iii) Approach to answer RQ2: Source Repositories Mining.

\subsection{Data Collection and Preparation}
This exploratory study focuses on the community-maintained ecosystem of Obsidian plugins. We retrieved the full registry\footnote{\url{https://github.com/obsidianmd/obsidian-releases/blob/master/community-plugins.json}} on October 20, 2025, obtaining 2{,}667 plugins with metadata such as name, author, description, and repository URL.
For a manageable yet representative analysis, we sampled 400 plugins. This sample size provides a 95\% confidence level with a 5\% margin of error for population proportions, establishing a robust and unbiased corpus for the subsequent clustering and topic modeling tasks

Because many plugin descriptions were too short to capture functionality, we enriched them by extracting additional keyphrases from each plugin’s README file. For every repository, all README files were collected and processed with the open-source LLM \textit{gpt-oss-120b}\footnote{\url{https://huggingface.co/unsloth/gpt-oss-120b-GGUF}} (MXFP4, 16-bit GGUF, served via \emph{Ollama}), which was prompted to generate two to four concise keyphrases summarizing core features.

To maintain linguistic consistency, non-English READMEs were automatically filtered out. Four repositories were removed during this step, resulting in 396 valid plugin instances used for subsequent analysis.

\subsection{Approach to answer RQ1: Topic Modeling and Labeling}
With the goal of identifying the main themes to which the plugins belong, we employed the \textit{BERTopic} library\footnote{\url{https://maartengr.github.io/BERTopic/index.html}}, a widely used framework for topic modeling and document clustering.  

The first step was to create a set of input documents. For each plugin, we produced one document by concatenating its name, short description, and extracted keyphrases. 
Vector embeddings were then generated using the \textit{Qwen/Qwen3-Embedding-0.6B}\footnote{\url{https://huggingface.co/Qwen/Qwen3-Embedding-0.6B-GGUF}} model (16-bit GGUF export, served through Ollama). The resulting embeddings were L2 normalized.\footnote{https://scikit-learn.org/stable/modules/generated/sklearn.preprocessing.normalize.html}

We applied the default \textit{BERTopic} pipeline, starting with the dimensionality reduction performed by the UMAP algorithm~\cite{mcinnes2020umapuniformmanifoldapproximation}, using as parameters $n\_neighbors=5$, $min\_dist=0.0$ and $metric=``cosine"$ to give more attention to the local structure.
Then a clustering phase was performed via \emph{HDBSCAN}~\cite{campello2013density}, and topic identification and representation was performed using Bag-of-Words models. This identifies a set of representative terms describing each topic.

\begin{figure*}[t]
	\centering
	\includegraphics[width=\linewidth]{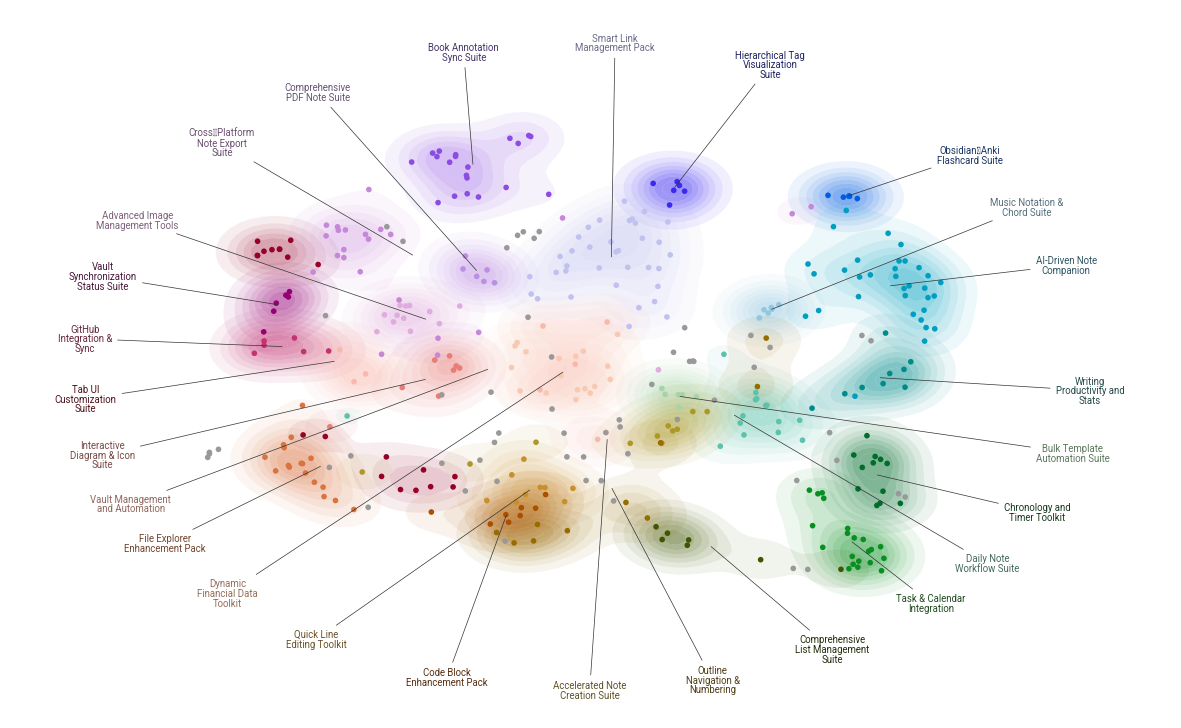}
	\caption{Documents distribution plot of the analyzed input documents}
	\label{fig:res_documents2}
\end{figure*}

The final phase is the LLM-based label generation, were we used \textit{gpt-oss-120b} (same setting as before).   
Iteratively, processing topics in a randomized sequence, the model received a random sample of five plugins (name, description, keyphrases, and topic terms) and was prompted to generate a concise title label. The previously assigned labels were also included in the prompt to avoid redundancy.

In addition, a second label has been generated, i.e., parent label, by grouping similar topic using the  the hierarchical topics function of \textit{BERTopic}\footnote{\url{https://maartengr.github.io/BERTopic/api/plotting/hierarchy.html}}, which basically builds a dendrogram and identifies the cutoff distance.
For each parent topic, we generated a label using the same procedure as before, considering each parent topic as a distinct cluster.


\subsection{Approach to answer RQ2: Source Repositories Mining}

\input{tables/tab-topics-count-colorbyparent}

To address RQ2, we mined the source software repositories for each plugin instance in our plugin list. Since all repositories were hosted on GitHub, we used \emph{PyGitHub}\footnote{https://github.com/PyGitHub/PyGitHub} to access the GitHub APIs. For each repository, we extracted the number of open, closed, and total pull requests. Then, for each topic, we aggregated this data by computing descriptive statistics (sum, mean, standard deviation, min, and max).

\begin{figure}
	\centering
	\includegraphics[width=1\columnwidth]{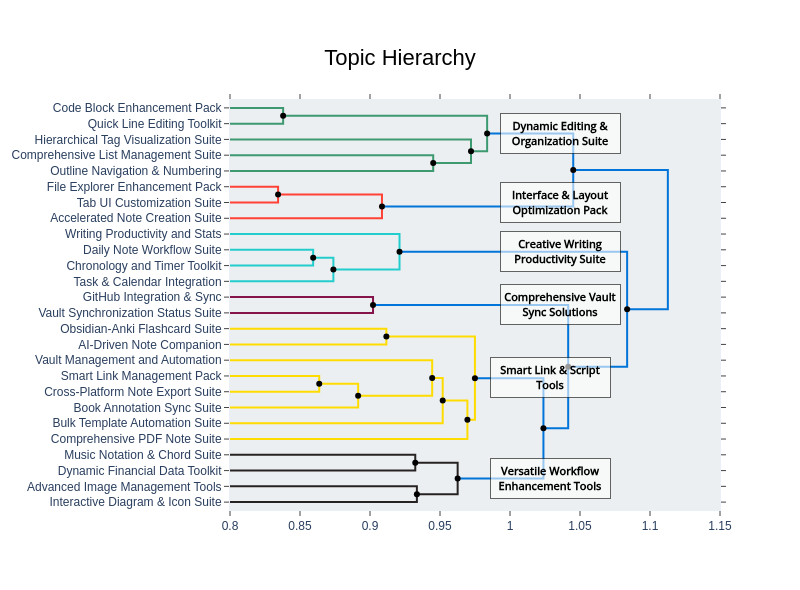}
	\caption{Dendrogram plot for topics hierarchy}
	\label{fig:res_dendrogram_custom_labels}
\end{figure}

\section{Preliminary Results}

This section summarizes the preliminary findings from our exploratory analysis of the \textit{Obsidian} plugin ecosystem, structured around two research questions.  

\subsection{Plugin Topics in the Obsidian Ecosystem}

Our analysis of 396 plugins shows that the Obsidian ecosystem is both diverse and semantically coherent, producing a total of 26 different topics, grouped in 6 different parent topics.
57 instances were not assigned to any of the identified topics.

Figure~\ref{fig:res_documents2} displays the two-dimensional projection of plugin embeddings, where clusters reveal distinct functional areas.
The most densely populated regions correspond to productivity-centric workflows—daily note taking, task and calendar integration, writing support, and accelerated note creation—highlighting Obsidian’s role as a workspace for knowledge-centric workflows rather than software development.

More specialized or creative functionalities appear in peripheral clusters, such as diagram editing, image manipulation, PDF handling, and music notation.  
Although smaller in size, these groups form cohesive semantic neighborhoods, showing how community developers extend Obsidian for diverse use cases within a knowledge-centric community.

The hierarchical perspective in Figure~\ref{fig:res_dendrogram_custom_labels} further illustrates how these topics relate at broader levels of abstraction.
We reported a parent labels to identify each major group of topics, defined by different colors.
Productivity-centric plugins (writing, tasks, templating) merge into a larger family of workflow-support tools.
Synchronization, exporting, and automation functionalities form a second domain centered on interoperability and vault management.  
Creative and visualization plugins appear as smaller but distinct branches.

\begin{tcolorbox}[colback=gray!5!white,colframe=gray,title=Answer to RQ1] We identified six types of plugins related to knowledge management and tooling. Our taxonomy highlighted six labels: (i) dynamic editing and organization, (ii) interface and layouts, (iii) creative writing and productivity, (iv) knowledge sync solutions, (v) linking and script tools, and (vi) workflow enhancements tools. 
\end{tcolorbox}

\subsection{Maintenance of the Plugins}

Table~\ref{tab:topics_pr_metrics} shows pull-request (PR) activity aggregated per topic. The results reveal substantial variation in maintenance intensity but overall consistent evidence of active development across the ecosystem. High-activity topics such as \textit{Vault Management and Automation}, \textit{Task \& Calendar Integration}, and \textit{Tab UI Customization Suite} stand out; in particular, Topic~3 (\textit{Task \& Calendar Integration}) has the highest total number of PRs and also the largest number of both open and closed PRs.

In contrast, more specific or creative categories—such as \textit{Music Notation \& Chord Suite} or \textit{Interactive Diagram \& Icon Suite}—show lower PR volumes, reflecting smaller user bases and narrower functional scope. Notably, Topic~14 (\textit{Quick Line Editing Toolkit}) records the lowest number of total PRs and also the minimum values across all related metrics.

Across nearly all topics, closed PRs far exceed open ones, indicating that contributors tend to process and resolve incoming work. Despite being used largely by non-developers, the ecosystem exhibits maintenance patterns similar to mature software environments, including iterative improvement, steady contribution flow, and topic-specific concentrations of activity.

\begin{tcolorbox}[colback=gray!5!white,colframe=gray,title=Answer to RQ2]
The results show substantial variation in maintenance intensity, but clear evidence of ongoing development with PRs receiving and closing Pull Requests.
The most actively maintained plugins belong to \textit{Task \& Calendar Integration} (Topic~3), which has the highest total number of PRs, the largest number of closed PRs, and the most open PRs.
At the other end of the spectrum, \textit{Quick Line Editing Toolkit} (Topic~14) shows the lowest levels of PR activity across all metrics.
\end{tcolorbox}

\section{Research Plan}

Building on these findings, our results suggest that non-developer ecosystems do experience software evolution. This motivates several focused research directions outlined below.

\subsection{Developers Intertwined with Knowledge-Centric }
Our analysis indicates active maintenance, suggesting that developer support may originate both inside and outside the Obsidian ecosystem. We aim to understand the sources and nature of this support through the following guiding questions:
\begin{itemize}
    \item Who maintains these plugins—contributors from within the ecosystem, external developers, or both—and how do they relate to the knowledge-centric community?
    \item What technical skills do maintainers possess, and to what extent do knowledge-centric communities depend on developer expertise for sustainability?
    \item Do knowledge-centric communities intersect with developer communities, and what forms do these cross-community interactions take?
\end{itemize}

\subsection{Deepening the Software Evolution Analysis}
Pull-request activity provides an initial view of sustainability, but hybrid ecosystems require richer indicators. We will therefore ask:
\begin{itemize}
    \item How do issue activity, commit patterns, contributor diversity, and release cadence vary across plugin groups?
    \item Do knowledge-centric plugins follow evolution and maintenance patterns comparable to developer-centric ecosystems?
\end{itemize}

\subsection{Knowledge-Centric vs.\ Developer Ecosystem Comparisons}
Although Obsidian shares structural traits with traditional ecosystems, the extent and meaning of this resemblance remain open questions:
\begin{itemize}
    \item How does Obsidian’s functional landscape compare with ecosystems such as VS Code, JetBrains Marketplace, or the Chrome Web Store?
    \item How does maintenance intensity differ between ecosystems created for knowledge work and those designed for software development?
\end{itemize}

This plan outlines the key steps required to expand the preliminary study into a full investigation of knowledge-centric plugin ecosystems.

\section{Potential Risks and Limitations}

As a research preview, this study presents early findings and therefore carries several limitations that will be addressed in the full version of the work.

First, our analysis is based on a representative subset of 396 plugins rather than the complete ecosystem of 2{,}667 plugins. While sufficient for exploratory insights, this sampling may overlook long-tail functionalities or subtle topic boundaries that only emerge at full scale. Extending the analysis to the full dataset is therefore a central objective of the next phase.

Second, the methodology relies on LLM-generated keyphrases and topic labels. Although effective for early analysis, these steps introduce potential biases related to prompting, sampling strategy, and model behaviour. 
The absence of a gold-standard taxonomy for knowledge-centric plugins also limits our ability to validate clusters quantitatively. To mitigate this, future work will verify the topic hierarchy consistency, along with cross-model comparisons and manual validation.

Third, the maintainability analysis currently uses pull-request activity as the sole indicator of sustainability. While PRs provide a useful initial signal, they capture only one dimension of ecosystem health. The full study will integrate additional repository-level metrics—including issues, commits, contributor activity, and release cadence—to develop a more comprehensive view of maintenance practices in this hybrid ecosystem.

Finally, Obsidian’s mixed community of developers and non-developers may follow contribution and maintenance patterns that differ from those observed in traditional developer-centric ecosystems. 
This makes Obsidian a valuable case for understanding how maintenance dynamics evolve in knowledge-centric plugin ecosystems.


\section{Conclusion}

This research preview presented an initial exploration of the Obsidian plugin ecosystem through the lens of software engineering. 
As output, we produced a first data-driven taxonomy, providing an early view of its maintenance characteristics that exhibits organizational patterns (e.g., functional domains, active pull-request activity) reminiscent of traditional software, suggesting developers are central to maintaining this hybrid, user-focused environment.

More broadly, the results show that modern software ecosystems increasingly operate at the intersection of coding and thinking. As platforms like Obsidian grow, the boundary between development tools and knowledge-centric tools becomes more fluid, raising new questions for software engineering about how such ecosystems form, evolve, and are maintained.
The next phase of this work will by to execute our research plan at scale.
Advancing this research agenda will contribute to a broader understanding of software ecosystems—one that spans programming practices, knowledge-centric workflows, and community-driven innovation.


\IEEEtriggeratref{43}
\bibliographystyle{IEEEtran}
\bibliography{IEEEabrv, references}

\end{document}

%% file: tables/tab-topics-count-colorbyparent.tex
\definecolor{parentDynamic}{RGB}{0,100,0}        
\definecolor{parentInterface}{RGB}{178,34,34}    
\definecolor{parentCreative}{RGB}{0,128,128}     
\definecolor{parentComprehensive}{RGB}{128,0,128}
\definecolor{parentSmart}{RGB}{184,134,11}       
\definecolor{parentVersatile}{RGB}{0,0,0}       

\begin{table*}[t]
\centering
\caption{Summary of Topics with Pull Request Metrics. 
\textcolor{blue!60!black}{Blue} indicates the maximum value in a column; 
\textcolor{red!60!black}{Red} indicates the minimum value.
Parent label colours match the cluster colours in Figure~\ref{fig:res_dendrogram_custom_labels}.}
\resizebox{\linewidth}{!}{%
\begin{tabular}{rlllrrrrrrrr}
\toprule
\textbf{Topic} & \textbf{Count} & \textbf{Topic Label} & \textbf{Parent Label} &
\shortstack{\textbf{PRs}\\\textbf{Sum}} &
\shortstack{\textbf{PRs}\\\textbf{Max}} &
\shortstack{\textbf{PRs}\\\textbf{Mean}} &
\shortstack{\textbf{PRs}\\\textbf{SD}} &
\shortstack{\textbf{Open}\\\textbf{Sum}} &
\shortstack{\textbf{Open}\\\textbf{Max}} &
\shortstack{\textbf{Closed}\\\textbf{Sum}} &
\shortstack{\textbf{Closed}\\\textbf{Max}} \\
\midrule
11 & 10 & Outline Navigation \& Numbering &
\textbf{\textcolor{parentDynamic}{Dynamic Editing \& Organization Suite}} &
49 & 30 & 4.90 & 9.29 & 4 & 3 & 45 & 29 \\
14 & 9 & Quick Line Editing Toolkit &
\textbf{\textcolor{parentDynamic}{Dynamic Editing \& Organization Suite}} &
\textbf{\textcolor{red!60!black}{3}} &
\textbf{\textcolor{red!60!black}{2}} &
\textbf{\textcolor{red!60!black}{0.33}} & 0.71 &
\textbf{\textcolor{red!60!black}{0}} &
\textbf{\textcolor{red!60!black}{0}} &
\textbf{\textcolor{red!60!black}{3}} &
\textbf{\textcolor{red!60!black}{2}} \\
17 & 8 & Code Block Enhancement Pack &
\textbf{\textcolor{parentDynamic}{Dynamic Editing \& Organization Suite}} &
75 & 20 & 9.38 & 14.01 & 2 & 1 & 73 & 19 \\
18 & 7 & Comprehensive List Management Suite &
\textbf{\textcolor{parentDynamic}{Dynamic Editing \& Organization Suite}} &
29 & 10 & 4.14 & 3.72 & 0 & 0 & 29 & 10 \\
20 & 6 & Hierarchical Tag Visualization Suite &
\textbf{\textcolor{parentDynamic}{Dynamic Editing \& Organization Suite}} &
49 & 22 & 8.17 & 7.78 & 0 & 0 & 49 & 22 \\
\midrule
4 & 20 & Tab UI Customization Suite &
\textbf{\textcolor{parentInterface}{Interface \& Layout Optimization Pack}} &
493 & \textbf{\textcolor{blue!60!black}{363}} & 24.65 & 80.63 &
16 & 4 & 477 & \textbf{\textcolor{blue!60!black}{359}} \\
7 & 18 & File Explorer Enhancement Pack &
\textbf{\textcolor{parentInterface}{Interface \& Layout Optimization Pack}} &
102 & 62 & 5.67 & 15.08 & 11 & 11 & 91 & 62 \\
12 & 10 & Accelerated Note Creation Suite &
\textbf{\textcolor{parentInterface}{Interface \& Layout Optimization Pack}} &
18 & 15 & 1.80 & 4.66 & 6 & 6 & 12 & 9 \\
\midrule
3 & 21 & Task \& Calendar Integration &
\textbf{\textcolor{parentCreative}{Creative Writing Productivity Suite}} &
\textbf{\textcolor{blue!60!black}{619}} & 182 & 29.48 & 49.26 &
\textbf{\textcolor{blue!60!black}{43}} &
\textbf{\textcolor{blue!60!black}{25}} &
\textbf{\textcolor{blue!60!black}{576}} & 179 \\
8 & 15 & Daily Note Workflow Suite &
\textbf{\textcolor{parentCreative}{Creative Writing Productivity Suite}} &
300 & 87 & 20.00 & 27.54 & 6 & 1 & 294 & 87 \\
10 & 12 & Chronology and Timer Toolkit &
\textbf{\textcolor{parentCreative}{Creative Writing Productivity Suite}} &
109 & 29 & 9.08 & 9.89 & 14 & 4 & 95 & 26 \\
13 & 9 & Writing Productivity and Stats &
\textbf{\textcolor{parentCreative}{Creative Writing Productivity Suite}} &
12 & 4 & 1.33 & 2.00 & 0 & 0 & 12 & 4 \\
\midrule
22 & 6 & Vault Synchronization Status Suite &
\textbf{\textcolor{parentComprehensive}{Comprehensive Vault Sync Solutions}} &
53 & 22 & 8.83 & 10.11 & 0 & 0 & 53 & 22 \\
23 & 6 & GitHub Integration \& Sync &
\textbf{\textcolor{parentComprehensive}{Comprehensive Vault Sync Solutions}} &
42 & 19 & 7.00 & 8.10 & 0 & 0 & 42 & 19 \\
\midrule
0 & 35 & Smart Link Management Pack &
\textbf{\textcolor{parentSmart}{Smart Link \& Script Tools}} &
268 & 97 & 7.66 & 18.19 & 36 & 7 & 232 & 92 \\
1 & 33 & AI-Driven Note Companion &
\textbf{\textcolor{parentSmart}{Smart Link \& Script Tools}} &
221 & 76 & 6.70 & 14.27 & 26 & 8 & 195 & 76 \\
2 & 23 & Cross-Platform Note Export Suite &
\textbf{\textcolor{parentSmart}{Smart Link \& Script Tools}} &
142 & 42 & 6.17 & 10.54 & 21 & 5 & 121 & 39 \\
5 & 20 & Book Annotation Sync Suite &
\textbf{\textcolor{parentSmart}{Smart Link \& Script Tools}} &
438 & 125 & 21.90 & 37.41 & 23 & 6 & 415 & 119 \\
16 & 8 & Vault Management and Automation &
\textbf{\textcolor{parentSmart}{Smart Link \& Script Tools}} &
351 & 305 & \textbf{\textcolor{blue!60!black}{43.88}} & 113.49 & 2 & 1 & 349 & 304 \\
19 & 6 & Comprehensive PDF Note Suite &
\textbf{\textcolor{parentSmart}{Smart Link \& Script Tools}} &
20 & 8 & 3.33 & 3.72 & 0 & 0 & 20 & 8 \\
21 & 6 & Bulk Template Automation Suite &
\textbf{\textcolor{parentSmart}{Smart Link \& Script Tools}} &
19 & 8 & 3.17 & 4.07 & 0 & 0 & 19 & 8 \\
25 & 6 & Obsidian-Anki Flashcard Suite &
\textbf{\textcolor{parentSmart}{Smart Link \& Script Tools}} &
31 & 12 & 5.17 & 6.94 & 0 & 0 & 31 & 12 \\
\midrule
6 & 18 & Dynamic Financial Data Toolkit &
\textbf{\textcolor{parentVersatile}{Versatile Workflow Enhancement Tools}} &
256 & 134 & 14.22 & 32.50 & 16 & 5 & 240 & 129 \\
9 & 13 & Advanced Image Management Tools &
\textbf{\textcolor{parentVersatile}{Versatile Workflow Enhancement Tools}} &
106 & 62 & 8.15 & 16.67 & 18 & 10 & 88 & 52 \\
15 & 8 & Interactive Diagram \& Icon Suite &
\textbf{\textcolor{parentVersatile}{Versatile Workflow Enhancement Tools}} &
48 & 12 & 6.00 & 4.93 & 1 & 1 & 47 & 12 \\
24 & 6 & Music Notation \& Chord Suite &
\textbf{\textcolor{parentVersatile}{Versatile Workflow Enhancement Tools}} &
25 & 18 & 4.17 & 6.55 & 0 & 0 & 25 & 18 \\
\bottomrule
\end{tabular}
}%
\label{tab:topics_pr_metrics}
\end{table*}

%% file: main.bbl
\begin{thebibliography}{10}
\providecommand{\url}[1]{#1}
\csname url@samestyle\endcsname
\providecommand{\newblock}{\relax}
\providecommand{\bibinfo}[2]{#2}
\providecommand{\BIBentrySTDinterwordspacing}{\spaceskip=0pt\relax}
\providecommand{\BIBentryALTinterwordstretchfactor}{4}
\providecommand{\BIBentryALTinterwordspacing}{\spaceskip=\fontdimen2\font plus
\BIBentryALTinterwordstretchfactor\fontdimen3\font minus \fontdimen4\font\relax}
\providecommand{\BIBforeignlanguage}[2]{{%
\expandafter\ifx\csname l@#1\endcsname\relax
\typeout{** WARNING: IEEEtran.bst: No hyphenation pattern has been}%
\typeout{** loaded for the language `#1'. Using the pattern for}%
\typeout{** the default language instead.}%
\else
\language=\csname l@#1\endcsname
\fi
#2}}
\providecommand{\BIBdecl}{\relax}
\BIBdecl

\bibitem{edirimannage2024developersvictimscomprehensive}
\BIBentryALTinterwordspacing
S.~Edirimannage, C.~Elvitigala, A.~K.~K. Don, W.~Daluwatta, P.~Wijesekara, and I.~Khalil, ``Developers are victims too : A comprehensive analysis of the vs code extension ecosystem,'' 2024. [Online]. Available: \url{https://arxiv.org/abs/2411.07479}
\BIBentrySTDinterwordspacing

\bibitem{Amershi2019}
\BIBentryALTinterwordspacing
S.~Amershi, D.~Weld, M.~Vorvoreanu, A.~Fourney, B.~Nushi, P.~Collisson, J.~Suh, S.~Iqbal, P.~N. Bennett, K.~Inkpen, J.~Teevan, R.~Kikin-Gil, and E.~Horvitz, ``Guidelines for human-ai interaction,'' in \emph{Proceedings of the 2019 CHI Conference on Human Factors in Computing Systems}, ser. CHI '19.\hskip 1em plus 0.5em minus 0.4em\relax New York, NY, USA: Association for Computing Machinery, 2019, p. 1–13. [Online]. Available: \url{https://doi.org/10.1145/3290605.3300233}
\BIBentrySTDinterwordspacing

\bibitem{Marron2024}
\BIBentryALTinterwordspacing
M.~Marron, ``A new generation of intelligent development environments,'' in \emph{Proceedings of the 1st ACM/IEEE Workshop on Integrated Development Environments}, ser. IDE '24.\hskip 1em plus 0.5em minus 0.4em\relax New York, NY, USA: Association for Computing Machinery, 2024, p. 43–46. [Online]. Available: \url{https://doi.org/10.1145/3643796.3648452}
\BIBentrySTDinterwordspacing

\bibitem{Hsu_2024}
\BIBentryALTinterwordspacing
S.~Hsu, M.~Tran, and A.~Fass, ``What is in the chrome web store?'' in \emph{Proceedings of the 19th ACM Asia Conference on Computer and Communications Security}, ser. ASIA CCS ’24.\hskip 1em plus 0.5em minus 0.4em\relax ACM, Jul. 2024, p. 785–798. [Online]. Available: \url{http://dx.doi.org/10.1145/3634737.3637636}
\BIBentrySTDinterwordspacing

\bibitem{linwordpress2023}
\BIBentryALTinterwordspacing
J.~Lin, M.~Sayagh, and A.~E. Hassan, ``The co-evolution of the wordpress platform and its plugins,'' \emph{ACM Trans. Softw. Eng. Methodol.}, vol.~32, no.~1, Feb. 2023. [Online]. Available: \url{https://doi.org/10.1145/3533700}
\BIBentrySTDinterwordspacing

\bibitem{zanartu2022automatically}
\BIBentryALTinterwordspacing
F.~Zanartu, C.~Treude, B.~Cartaxo, H.~S. Borges, P.~Moura, M.~Wagner, and G.~Pinto, ``Automatically categorising github repositories by application domain,'' 2022. [Online]. Available: \url{https://arxiv.org/abs/2208.00269}
\BIBentrySTDinterwordspacing

\bibitem{sas2023gitranking}
C.~Sas, A.~Capiluppi, C.~Sipio, J.~Rocco, and D.~\{Di Ruscio\}, ``\BIBforeignlanguage{English}{Gitranking: A ranking of github topics for software classification using active sampling},'' \emph{\BIBforeignlanguage{English}{Software - Practice and Experience}}, vol.~53, no.~10, pp. 1982--2006, Oct. 2023, publisher Copyright: {\textcopyright} 2023 The Authors. Software: Practice and Experience published by John Wiley \& Sons Ltd.

\bibitem{nguyen2024gavel}
P.~Nguyen, J.~Rocco, C.~Di~Sipio, M.~Shakya, D.~Di~Ruscio, and M.~Di~Penta, ``Automatic categorization of github actions with transformers and few-shot learning,'' 07 2024.

\bibitem{Sergeyuk2024}
\BIBentryALTinterwordspacing
A.~Sergeyuk, S.~Titov, and M.~Izadi, ``In-ide human-ai experience in the era of large language models; a literature review,'' in \emph{Proceedings of the 1st ACM/IEEE Workshop on Integrated Development Environments}, ser. IDE '24.\hskip 1em plus 0.5em minus 0.4em\relax New York, NY, USA: Association for Computing Machinery, 2024, p. 95–100. [Online]. Available: \url{https://doi.org/10.1145/3643796.3648463}
\BIBentrySTDinterwordspacing

\bibitem{Nghiem2024}
\BIBentryALTinterwordspacing
K.~Nghiem, A.~M. Nguyen, and N.~Bui, ``Envisioning the next-generation ai coding assistants: Insights \& proposals,'' in \emph{Proceedings of the 1st ACM/IEEE Workshop on Integrated Development Environments}, ser. IDE '24.\hskip 1em plus 0.5em minus 0.4em\relax New York, NY, USA: Association for Computing Machinery, 2024, p. 115–117. [Online]. Available: \url{https://doi.org/10.1145/3643796.3648467}
\BIBentrySTDinterwordspacing

\bibitem{Ahmed2025}
\BIBentryALTinterwordspacing
T.~Ahmed, P.~Devanbu, C.~Treude, and M.~Pradel, ``Can llms replace manual annotation of software engineering artifacts?'' in \emph{2025 IEEE/ACM 22nd International Conference on Mining Software Repositories (MSR)}.\hskip 1em plus 0.5em minus 0.4em\relax IEEE, Apr. 2025, p. 526–538. [Online]. Available: \url{http://dx.doi.org/10.1109/MSR66628.2025.00086}
\BIBentrySTDinterwordspacing

\bibitem{Crupi2025}
G.~Crupi, R.~Tufano, A.~Velasco, A.~Mastropaolo, D.~Poshyvanyk, and G.~Bavota, ``On the effectiveness of llm-as-a-judge for code generation and summarization,'' \emph{IEEE Transactions on Software Engineering}, vol.~51, no.~8, pp. 2329--2345, 2025.

\bibitem{Gonzalez-Barahona2024}
J.~M. Gonzalez-Barahona, ``Software development in the age of llms and xr,'' in \emph{2024 IEEE/ACM First IDE Workshop (IDE)}, 2024, pp. 66--69.

\bibitem{qin2021chrome}
Z.~Qin, H.~Zhuang, R.~Jagerman, X.~Qian, P.~Hu, C.~Chen, X.~Wang, M.~Bendersky, and M.~Najork, ``Bootstrapping recommendations at chrome web store,'' 2021.

\bibitem{mcinnes2020umapuniformmanifoldapproximation}
\BIBentryALTinterwordspacing
L.~McInnes, J.~Healy, and J.~Melville, ``Umap: Uniform manifold approximation and projection for dimension reduction,'' 2020. [Online]. Available: \url{https://arxiv.org/abs/1802.03426}
\BIBentrySTDinterwordspacing

\bibitem{campello2013density}
R.~J. Campello, D.~Moulavi, and J.~Sander, ``Density-based clustering based on hierarchical density estimates,'' in \emph{Pacific-Asia conference on knowledge discovery and data mining}.\hskip 1em plus 0.5em minus 0.4em\relax Springer, 2013, pp. 160--172.

\end{thebibliography}
